%% file: main.tex
\documentclass[10pt]{article}

\input{preamble.tex}

\title{\sffamily\textbf{Object independent scatter sensitivities for PET, applied to scatter estimation through fast Monte Carlo simulation}}

\author[1,*]{Simon~No\"e~\orcidlink{0000-0003-0628-0130}}
\author[1]{Seyed~Amir~Zaman~Pour~\orcidlink{0000-0002-8816-9358}}
\author[1]{Ahmadreza~Rezaei~\orcidlink{0000-0002-9843-8080}}
\author[2]{Charles~Stearns~\orcidlink{0000-0002-7074-6663}}
\author[1]{Johan~Nuyts~\orcidlink{0000-0003-1923-6198}}
\author[1]{Georg~Schramm~\orcidlink{0000-0002-2251-3195}}

\affil[1]{Department of Imaging and Pathology, Division of Nuclear Medicine, KU Leuven}
\affil[2]{GE HealthCare, Waukesha, WI, United States.}
\affil[*]{Corresponding author: \href{mailto:simon.noe@kuleuven.be}{simon.noe@kuleuven.be}}

\begin{document}

\maketitle

\begin{abstract}
\input{abstract.tex}
\end{abstract}

\input{introduction.tex}

\input{methods.tex}

\input{results.tex}

\input{discussion.tex}

\input{conclusion.tex}


\footnotesize
\setlength{\itemsep}{0pt}
\setlength{\parskip}{0pt}
\bibliography{main.bib}
\bibliographystyle{ieeetr}

\end{document}

%% file: preamble.tex
\usepackage[a4paper,left=3cm,right=3cm,top=2cm,bottom=2cm]{geometry}
\usepackage[hidelinks,colorlinks=false]{hyperref}
\usepackage{authblk}

\usepackage{orcidlink}

\usepackage{amsmath}
\usepackage{amssymb}
\usepackage{amsfonts}
\usepackage{caption}
\usepackage{subcaption}
\usepackage{cleveref}
\usepackage{setspace}
\usepackage{afterpage}
\usepackage[final,nopatch=footnote]{microtype}
\usepackage{multirow}
\usepackage{booktabs}

\usepackage{sectsty}
\allsectionsfont{\sffamily}

%% file: abstract.tex
\textbf{Objective}
Scattered coincidences are a major source of quantitative bias in positron emission tomography (PET) and must be compensated during reconstruction using an estimate of scattered coincidences per line-of-response and time-of-flight bin.
Such estimates are typically obtained from simulators with simple cylindrical scanner models that omit detector physics.
Incorporating detector sensitivities for scatter is challenging, as scattered coincidences have less constrained properties (e.g., incidence angles) than true coincidences.

\textbf{Approach}
We integrated a 5D single-photon detection probability lookup table (photon energy, incidence angle, detector location) into the simulator logic.
The resulting scatter sinogram is multiplied by a precomputed, LUT-specific scatter sensitivity sinogram to yield the scatter estimate.
Scatter was simulated with MCGPU-PET, a fast Monte Carlo simulator with a simplified scanner model, and applied to phantom data from a simulated GE Signa PET/MR in GATE.
We evaluated three scenarios:
\begin{enumerate}
    \item Long, high-count MCGPU-PET simulations from a known activity distribution (reference).
    \item Same distribution with limited simulation time and counts.
    \item Same low-count data with joint estimation of activity and scatter during reconstruction.
\end{enumerate}
We also adapted the approach to test it on two acquisitions from a real Signa PET/MR.

\textbf{Main result}
In scenario 1, scatter-compensated reconstructions achieved $<1\%$ global bias in all active regions relative to true-only reconstructions.
In scenario 2, noisy scatter estimates caused strong positive bias, but Gaussian smoothing restored accuracy to scenario 1 levels.
In scenario 3, joint estimation under low-count conditions maintained $<1\%$ global bias in nearly all regions.
For real scans, the Monte Carlo-based scatter estimate was very similar to the vendor scatter estimate.

\textbf{Significance}
Although demonstrated with a fast Monte Carlo simulator, the proposed scatter sensitivity modeling could enhance existing single scatter simulators used clinically, which typically neglect detector physics.
This proof-of-concept also supports the feasibility of scatter estimation for real scans using fast Monte Carlo simulation, offering potentially greater accuracy and robustness to acquisition noise.

%% file: introduction.tex
\section{Introduction}
If not accounted for correctly, coincidences originating from scattered photons are a large source of quantitative bias in positron emission tomography (PET), as they represent 30-50\% of all non-random coincidences.
Properly compensating for scatter coincidences requires an estimate of the expected number of scatter coincidences for each time-of-flight (TOF) bin along each line-of-response (LOR).

In most cases (barring energy-based \cite{Hamill2024,Efthimiou2022} and some deep learning-based \cite{Qian2017} methods), the scatter estimate is calculated iteratively, by reconstructing an estimate of the true activity distribution with the current scatter estimate, simulating a new scatter estimate with this distribution, and then repeating this process a few times.
In clinical practice, the scatter estimates are simulated using tail-fitted single scatter simulation (SSS).
This procedure makes use of a few shortcuts to ensure the computation can be done within a time frame that is acceptable for routine clinical use, while still producing good results in almost all cases \cite{Ollinger1996,Watson2007,JosSanto2025}.
Monte Carlo (MC) simulation is an alternative to tail-fitted SSS that is believed to be more accurate and more robust to noise in the acquisition sinogram, due to the fact that it models most of the physical processes that take place during a scan.
MC Simulators like Gate \cite{Jan2004} and SimSet \cite{Harrison2012} can simulate very accurate scatter estimates \cite{Bayerlein2024}, but require a lot of computation time to do so, and thus are not feasible for use in clinical practice.
Recently though, very fast MC simulators were developed \cite{Herraiz2024,Ma2020,Scheins2021,Lai2019,Galve2024,Cabello2025}, and their feasibility for scatter estimation within clinically acceptable times was demonstrated \cite{No2024}.

In order to achieve fast simulation times, many of the aforementioned single scatter and MC simulators use a simple cylindrical geometry and do not simulate detector physics.
All photons that pass through the ``detector'' are kept, which also means that the probability of detection is assumed to be the same for all photons.
This is not what happens in real scanners, as evidenced by the sensitivity\footnote{By \textit{sensitivity sinogram}, we are referring to a correction sinogram which is \textit{multiplied} with another sinogram (e.g. the forward projection of the activity image in OSEM reconstruction) in order to incorporate the detector sensitivities of the scanner. \textit{Normalization} involves the removal of these detector sensitivities, by dividing by the sensitivity sinogram, or equivalently, by multiplying with the reciprocal of the sensitivity sinogram (sometimes called a \textit{normalization sinogram}).}
sinogram for true coincidences\footnote{A true coincidence denotes a pair of photons originating from the same annihilation, which reach the detectors without any interaction in the scanned object. For a scattered coincidence, at least one of the photons must undergo scatter inside the scanned object.} which vendors estimate for their scanners.
This sinogram contains structure resulting from the influence of scanner geometry (among others) on detection probability.
Multiplication with this sinogram is sufficient to incorporate the scanner's detector sensitivities in a forward projection of the activity image, but it would not work as well for a scatter sinogram returned by a cylindrical simulator that does not simulate detector physics.
The reason is that scattered photons have energies lower than 511 keV, and an incidence angle that may differ a lot from the one of unscattered photons along the same LOR, resulting in a difference in probability of detection for scattered and true coincidences, even along the same LOR.

\begin{figure}
    \centering
    \includegraphics[width=0.75\linewidth]{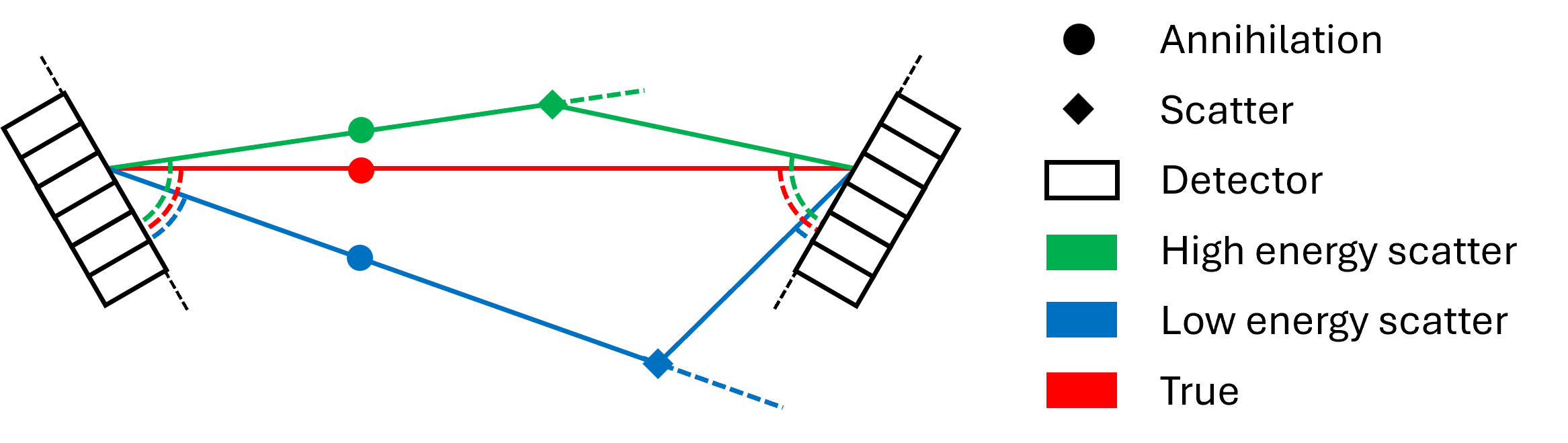}
    \caption{This schematic shows the difference in possible photon incidence angles for the same LOR. True coincidences (red) have a very limited distribution, as the two photons always travel along a straight line through the annihilation point. Scattered coincidences (green and blue) can connect the two detectors with a much larger degree of freedom, resulting in a larger possible distribution for the angles of incidence. The shape of this distribution is also correlated with the remaining photon energy.}
    \label{fig:incidence_schematic}
\end{figure}

In practice, scatter estimate sinograms from the simulator can be multiplied with a separate ``scatter sensitivity sinogram'' to account for this difference in sensitivity between trues and scatters \cite{Watson2007}.
While this sensitivity sinogram can compensate for overall differences in sensitivity between trues and scatters for each LOR, using the same scatter sensitivity for all objects assumes that the overall sensitivity difference between trues and scatters in each LOR is independent of the shape and size of the object in the scanner/simulator (similar to the trues sensitivity sinogram \cite{Badawi1998,Badawi1999}), which we will show is an inaccurate assumption.
For example, one component in the trues sensitivity sinogram is the intrinsic angle that any LOR makes to both detector faces, which can be taken into account using a single value for each LOR (i.e. a sinogram) for trues, because the incidence angle of a true photon will always be very close to the angle of the LOR to the detector face (Figure \ref{fig:incidence_schematic}).
For scattered photons, the incidence angle of the photon does not have to be the same as the angle of the LOR to the detector face at all, and will be correlated with the remaining energy of the photon after scattering.
The distribution of photon incidence angles and energies (and as a result, the scatter sensitivity of an LOR) depends heavily on the size and shape of the object in the scanner.
Therefore, the use of a scatter sensitivity sinogram alone is not appropriate, since it cannot account for the object dependence of the scatter sensitivity.
For scatter estimate sinograms from simulators that do not simulate detector physics, the multiplication with the scatter sensitivity sinogram should be supplemented with a sensitivity correction on the level of individual photons to achieve object independence.

In this work, we aim to demonstrate the following:
\begin{enumerate}
    \item The probability of detection varies significantly for single photons with different energies, angles of incidence and detector location in the scanner geometry.
    \item Compensating for the effects these probability differences have on scattered coincidences using only a scatter sensitivity sinogram can result in reconstruction artifacts due to its object dependence.
    \item Re-estimating the scatter sensitivity sinogram after adding a lookup table (LUT) to the scatter simulator logic (containing the probability of detection for single photons based on their characteristics) makes the sensitivity sinogram (nearly) object independent.
    \item This approach can be applied to fast, low-count MC simulations and still produce accurate scatter estimates, even when the activity distribution is unknown.
    \item Given some additional modifications to the simulator, the same fast, low-count MC simulations can be used to estimate scatter for real scans, without the need for tail fitting.
\end{enumerate}

We used the MCGPU-PET simulator \cite{Herraiz2024} as an example to demonstrate the feasibility of the LUT approach for any cylindrical simulator that does not simulate detector physics.
The resulting scatter sinograms were used to compensate scatter during reconstruction for three different phantoms on a replica of the Signa PET/MR, simulated in GATE. We evaluated the effectiveness of the scatter estimation procedures directly on the scatter-compensated reconstructions, as the accuracy of the scatter sinograms themselves has little practical relevance.
The evaluation was conducted for three scenarios:
\begin{enumerate}
    \item Known activity distribution, unlimited simulation time: Is it possible to get reconstruction accuracy close to a reconstruction of only trues in this ideal scenario?
    \item Known activity distribution, very limited simulation time: Is it possible to maintain the reconstruction accuracy of scenario 1 if the limited simulation time results in very noisy scatter estimates?
    \item Unknown activity distribution, very limited simulation time: Is it possible to maintain the reconstruction accuracy of scenario 1 when the same low-count simulations from scenario 2 are used in a more realistic scenario which requires the joint estimation of activity and scatter?
\end{enumerate}
After the quantitative evaluation on simulated acquisitions, we included the effect of inter-crystal scatter recovery, a correction factor for non-ideal detectors and the dead-time and pile-up corrections in the scatter estimation pipeline. We then estimated the scatter for two phantom acquisitions on a real Signa PET/MR and compared the resulting sinograms and reconstructions to those obtained using tail-fitted SSS from the vendor.

%% file: methods.tex
\section{Materials and Methods}

\subsection{Simulated scans in GATE}
\begin{figure}
    \centering
    \includegraphics[width=0.75\linewidth]{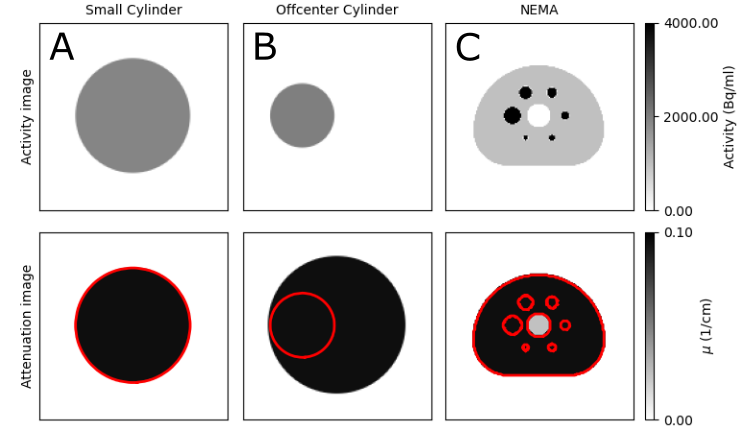}
    \caption{Voxelized representations of activity and attenuation for all 3 phantoms used in this study. The outline of the activity images is shown in red on the attenuation images.}
    \label{fig:phantoms}
\end{figure}

We first evaluated our approach on acquisitions using a replica of the GE Signa PET-MR scanner \cite{Levin2016} built in GATE, in order to have control over all scanner parameters, and to have access to the true activity images and coincidences.
In total, acquisitions were simulated for 3 different phantoms (Figure~\ref{fig:phantoms}).
In these simulated scans, we used 11.2\% as the energy resolution, and 425-650 keV as the energy window.
TOF resolution was set to 380 ps FWHM, similar to the real scanner.
Since accurate methods for randoms estimation and compensation are available and widely used, randoms were not simulated for our evaluation, only trues and scatters.
The effect of decay on activity concentration was not simulated.

\subsubsection{Signa PET/MR description}
The Signa PET/MR scanner \cite{Levin2016} is made up of 28 modules arranged in a ring configuration, with a small gap between the modules.
Each module is 16 crystals wide transaxially, resulting in a total of 448 crystals per detector ring.
Axially, each module consists of 5 detector blocks with a small gap between each block.
These blocks are 16 crystals wide and 9 crystals long, meaning the scanner has a total of 45 detector rings.
Each detector block consists of 4 detector units, which are 4 crystals wide and 9 crystals long, and delineate the area in which photon interactions are summed to produce singles. The scanner is 250.4 mm long, and the distance from the module face to the center of the scanner is 311.8 mm. We will refer to this last distance as the radius of the scanner.

\subsubsection{Small cylinder acquisition}
The first phantom (Figure \ref{fig:phantoms}A) consisted of a 250 mm diameter cylinder with the same length as the simulated scanner (250.4 mm).
The cylinder was filled with water, and a homogeneous concentration of 1.92 kBq/ml of F-18.
The total simulated scan time was 1 hour, resulting in $1.22 \cdot 10^9$ total coincidences and a scatter fraction of 35.2\%.

\subsubsection{Offcenter cylinder acquisition}
The second phantom (Figure \ref{fig:phantoms}B) was selected to create a challenge for scatter estimation \cite{Watson2020}.
It consisted of a small, hot cylinder (14 cm diameter; 2 kBq/ml), inside a larger, cold cylinder (30 cm diameter) filled with water.
The large cylinder was put in the center of the simulated scanner, while the small cylinder was offset to the left by 7.5 cm (but still inside the larger cylinder).
Both cylinders had the same length as the scanner.
The total simulated scan time was 3 hours, resulting in $7.84 \cdot 10^8$ total coincidences and a scatter fraction of 43.6\%.

\subsubsection{NEMA phantom acquisition}
For the third phantom (Figure \ref{fig:phantoms}C), we built a replica of the NEMA image quality phantom in GATE.
The spheres were filled with 4 kBq/ml and the background with 1 kBq/ml.
The total simulated scan time was 1 hour, resulting in $8.48 \cdot 10^8$ total coincidences and a scatter fraction of 33.8\%.

\subsubsection{Trues sensitivity sinogram}
In order to incorporate the detector sensitivities during reconstruction, a trues sensitivity sinogram was computed by dividing a GATE simulation sinogram containing only trues by a forward projection sinogram of the same phantom.
This phantom was a 602.4 mm diameter cylinder with the same axial extent (250.4 mm) as the simulated scanner, in order to cover the entire field-of-view (FOV) of the sinogram.
The effects of attenuation and decay were not simulated. With a concentration of 1 Bq/ml and a total simulated scan time of 26 hours, $6.04 \cdot 10^8$ true coincidences were obtained.
Radial symmetries were exploited to reduce the amount of noise in the sinogram.
The same trues sensitivity sinogram was used for all reconstructions.

\subsection{Scatter simulation}
\begin{figure}
    \centering
    \includegraphics[width=1.0\linewidth]{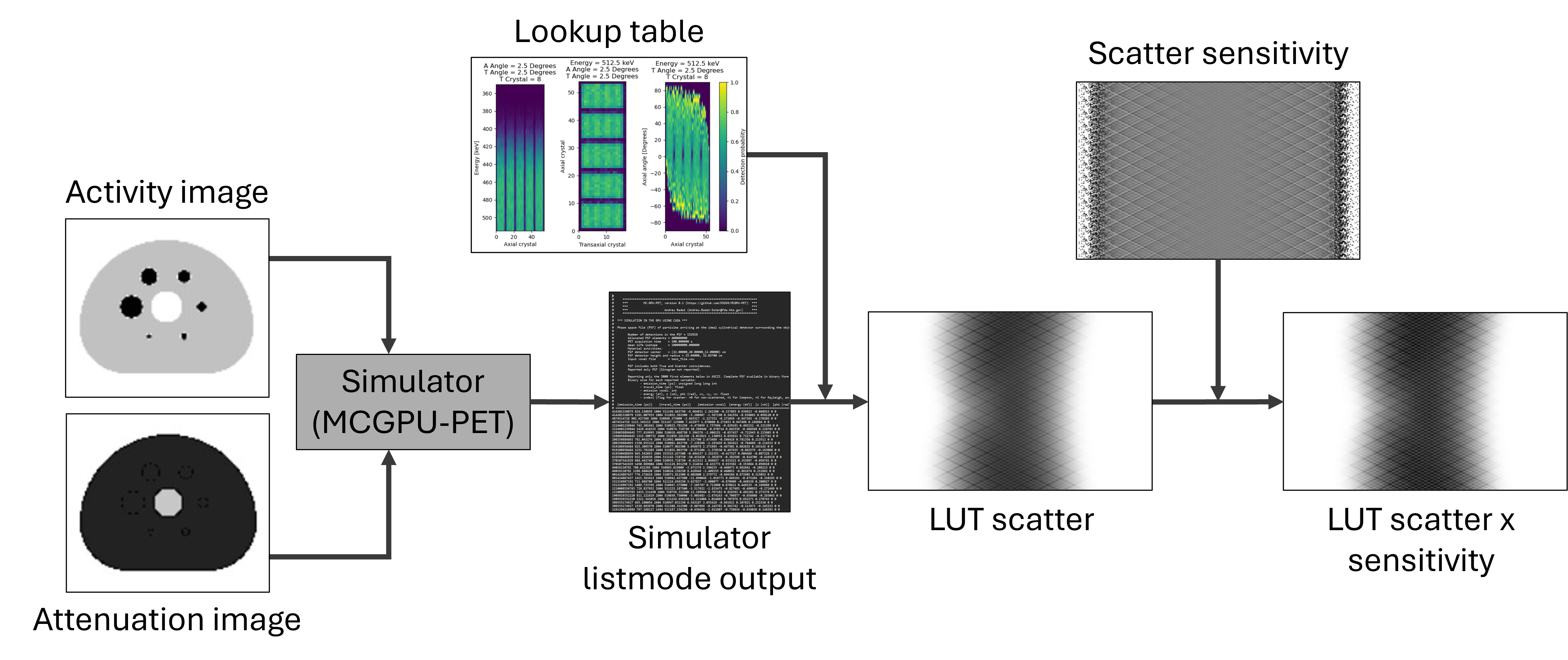}
    \caption{Schematic overview of our approach to scatter simulation. The simulator requires an activity and attenuation image as inputs, and outputs a list of coincidences. These coincidences are unlisted into an output sinogram which takes the photon detection probability LUT into account (LUT scatter), which is multiplied with the corresponding scatter sensitivity sinogram to produce the final scatter estimate (LUT scatter $\times$ sensitivity).}
    \label{fig:schematic_sensitivities}
\end{figure}

Our approach to the incorporation of scatter sensitivity in the scatter simulator output consists of two components: a LUT which assigns a probability of detection between 0 and 1 to every detected photon based on its properties, and a scatter sensitivity sinogram, which was calculated by dividing a GATE simulated scatter sinogram by an MCGPU-PET simulated scatter sinogram (with the LUT taken into account).
The exact derivation of this LUT is explained below.
The LUT is used during the unlisting of the MCGPU-PET listmode output into a sinogram, while the scatter sensitivity sinogram is multiplied with the sinogram resulting from the processed listmode data (Figure \ref{fig:schematic_sensitivities}).

\subsubsection{Scatter simulator description}
\begin{figure*}[t!]
    \centering
    \begin{subfigure}[t]{0.5\textwidth}
        \centering
        \includegraphics[height=1.2in]{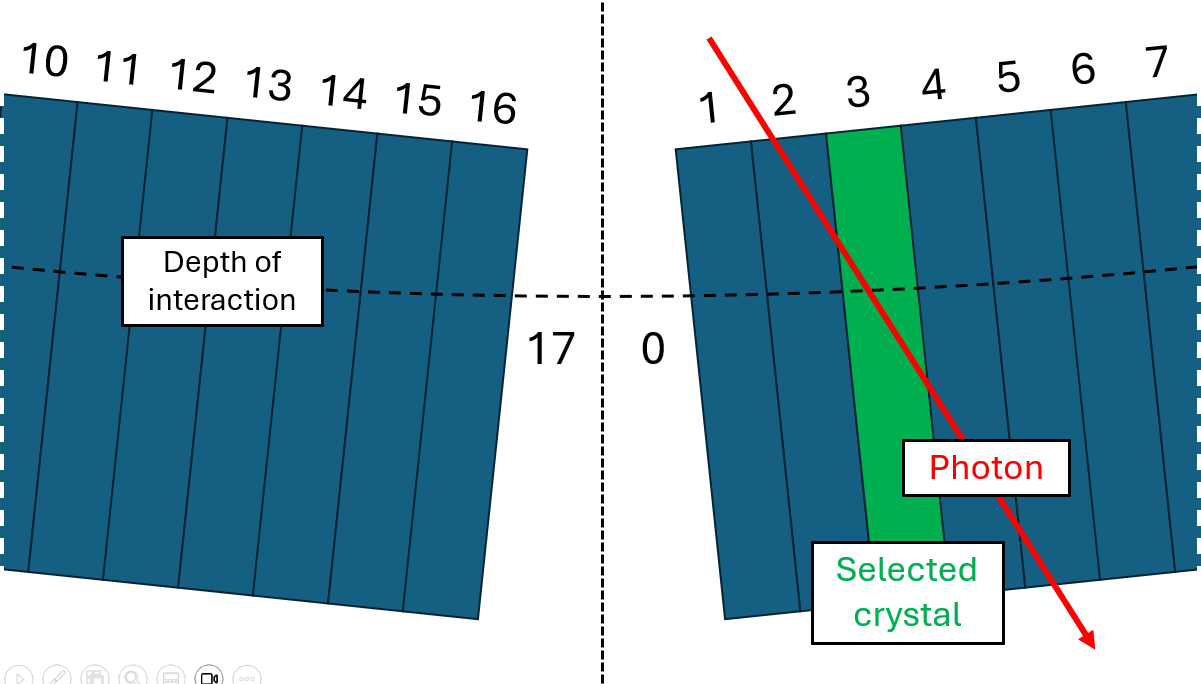}
        \caption{Transaxial section of the scanner.}
        \label{fig:crystals1}
    \end{subfigure}%
    ~ 
    \begin{subfigure}[t]{0.5\textwidth}
        \centering
        \includegraphics[height=1.2in]{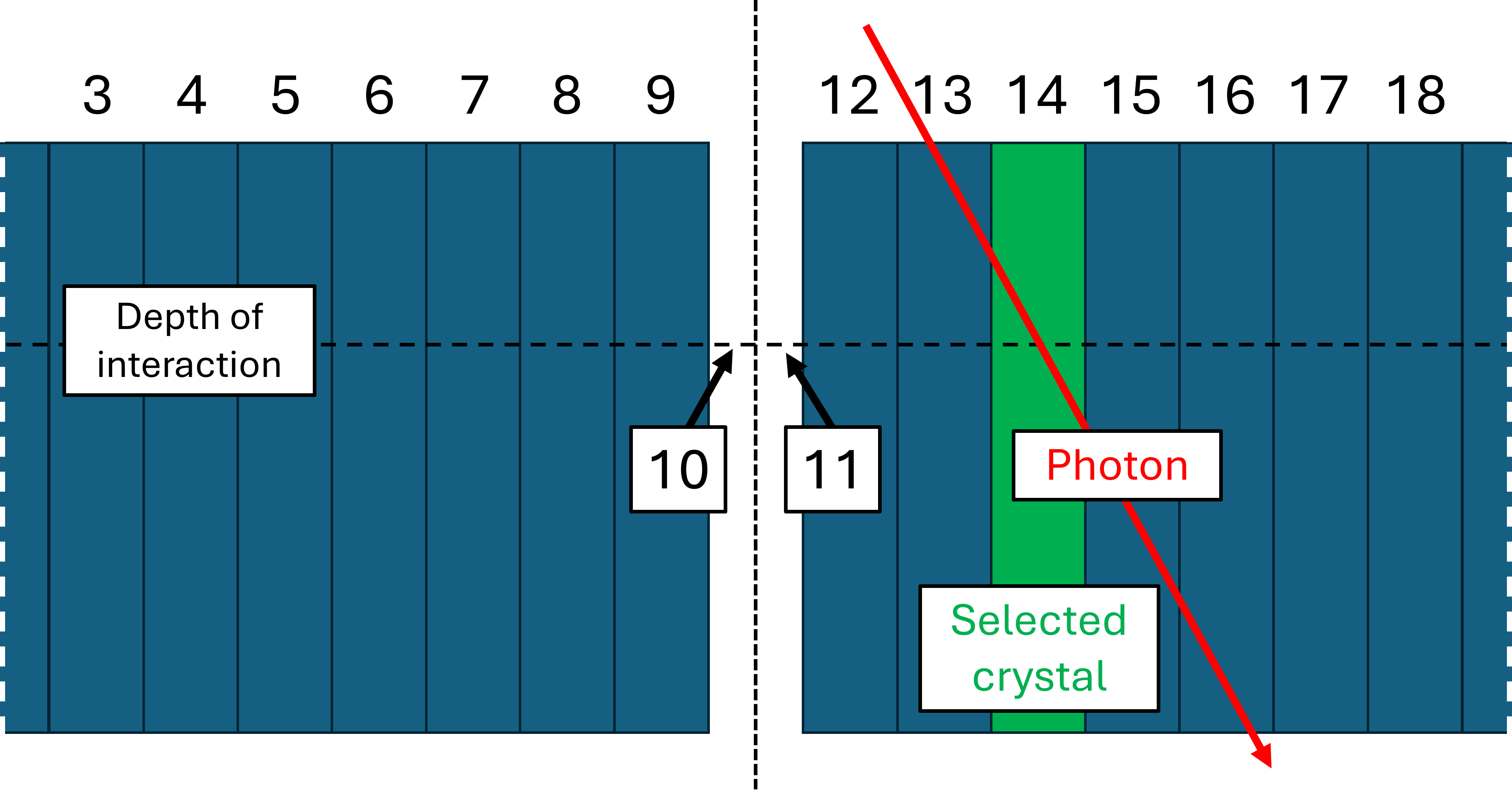}
        \caption{Axial section of the bottom of the scanner.}
        \label{fig:crystals2}
    \end{subfigure}
    \caption{Crystal definitions for MCGPU-PET and the detection probability LUT. The crystal that a photon is assigned to is selected based on the location of the intersection of the photon with the MCGPU-PET detector cylinder, located at the depth of interaction. Both the axial and transaxial configurations include virtual crystals, one for each side of the gaps between blocks.}
    \label{fig:crystals}
\end{figure*}

MCGPU-PET is a very fast GPU based simulator \cite{Herraiz2024} that uses a simple cylinder as the scanner geometry, and detects all photons that cross the cylinder surface (i.e. does not simulate detector interactions).
The simulator can produce a sinogram output, but our approach requires processing of the individual detected photons, which was done using the listmode output of the simulator.
For the simulator geometry, we used the same scanner length as in GATE (250.4 mm), but added the average depth of interaction (DOI; proprietary value) to the GATE scanner radius (311.8 mm), to better reflect the situation in the GATE simulations, where (on average) photons are detected at the DOI.
However, since the GATE scanner is a polygon and the MCGPU-PET detector is cylindrical, the simulated DOI is slightly shallower towards the edges of the transaxial modules (see Figure \ref{fig:crystals}\subref{fig:crystals1}).
When using the LUT, a near-perfect energy resolution of 0.1\% and an energy window between 300 and 650 keV were used.
When the LUT was not used, the same energy parameters as in GATE were used (11.2\% resolution and window between 425 and 650 keV).

After every simulation, MCGPU-PET outputs a list containing only non-random coincidences.
This list contains the emission time, emission voxel, photon energies, the location on the detector cylinder where each photon hit (in cylindrical coordinates), the direction that the photons were traveling in, and whether or not they underwent single, multiple, and/or Rayleigh scatter on their way to the detector.
The sinogram TOF bin was calculated from the difference in emission time between two coincident photons.
Using the position of detection, an axial and transaxial crystal position could be assigned to each photon (figure \ref{fig:crystals}), which was then used to put each coincidence in the corresponding sinogram bin.
Gaps in the scanner geometry were defined based on the solid angle they occupy (for transaxial gaps) and their width (for axial gaps) in the real scanner.
Photons hitting these gaps were either discarded or added to crystals at the block edges, depending on which LUT was used.
By default, every coincidence adds one count to the sinogram.
When using the LUT however, the value added to the sinogram was equal to the product of the detection probabilities of the two coincident photons.

\subsubsection{Photon detection probability LUT}
In order to compute a LUT with detection probabilities for each photon, it is necessary to know, for a photon detected with certain properties in MCGPU-PET, what the probability is that it would be detected in GATE.
The LUT was calculated based on a simulation of the same cylinder as used for the trues normalization (covering the entire FOV), with the difference that for this simulation, it only emitted single photons using a uniform energy spectrum between 0 and 511 keV, to cover all possible photon properties more uniformly.

\begin{figure*}[t!]
    \centering
    \begin{subfigure}[t]{0.5\textwidth}
        \centering
        \includegraphics[height=1.2in]{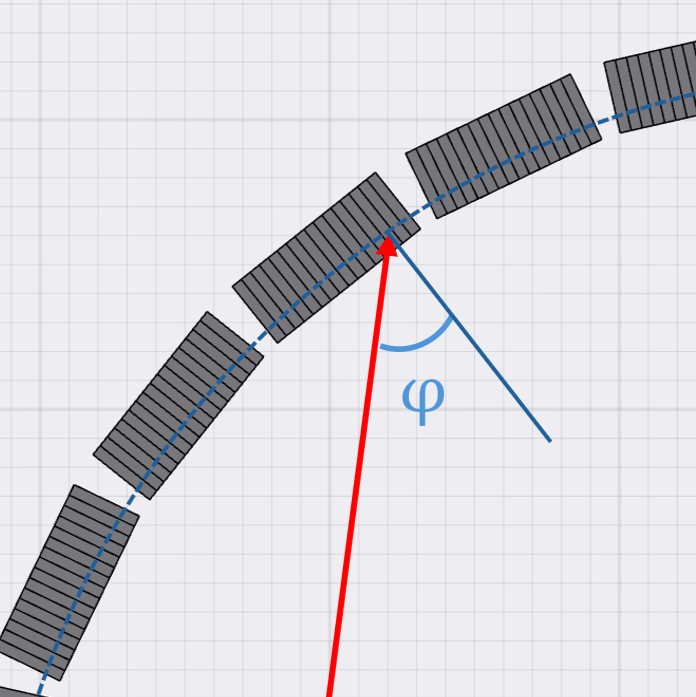}
        \caption{Transaxial section of the scanner.}
        \label{fig:angles1}
    \end{subfigure}%
    ~ 
    \begin{subfigure}[t]{0.5\textwidth}
        \centering
        \includegraphics[height=1.2in]{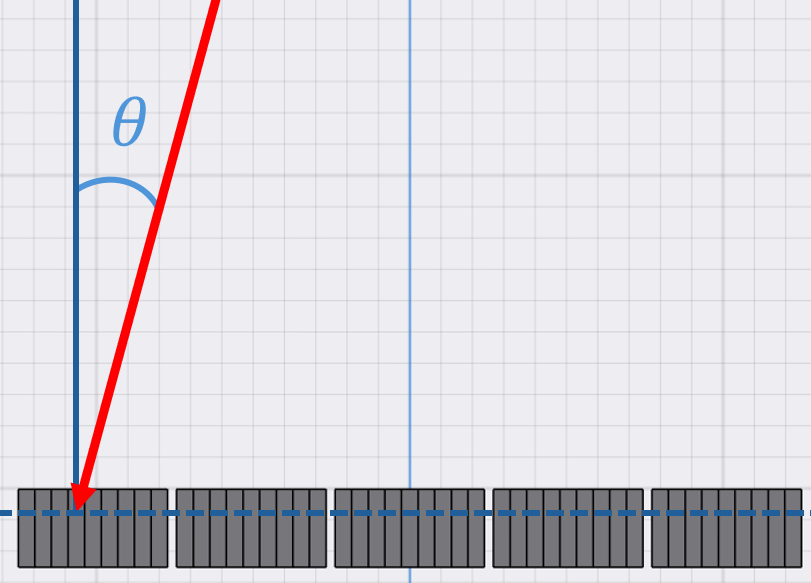}
        \caption{Axial section of the bottom of the scanner.}
        \label{fig:angles2}
    \end{subfigure}
    \caption{Angular definitions for the detection probability LUT. The (trans)axial angle is defined as the angle that the photon path makes in the (trans)axial plane with the position vector of its intersection with the detector cylinder. The range of both angles is [-90,90] degrees.}
    \label{fig:angles}
\end{figure*}

Detecting photons in a similar way to MCGPU-PET can be done in GATE by placing an ``Actor'' cylinder in front of the detector crystals, which detects any photon passing through it, and logs certain properties of the photon, such as its position, its event ID, its energy and the direction it was traveling in.
Photons were binned into a 5D histogram based on the following properties: the exact energy of the incoming photon (no energy resolution modeling), the angle the photon path forms with the position vector of the photon in the transaxial plane (transaxial angle, figure \ref{fig:angles}\subref{fig:angles1}), the angle the photon path forms with the position vector of the photon in the axial plane (axial angle, figure \ref{fig:angles}\subref{fig:angles2}), the transaxial location of the crystal in the module that the photon would be assigned to during unlisting of the MCGPU-PET output (figure \ref{fig:crystals}\subref{fig:crystals1}), and the axial location of the crystal that the photon would be assigned to during unlisting of the MCGPU-PET output (figure \ref{fig:crystals}\subref{fig:crystals2}).
The binsizes were 5 keV for photon energy, 18 indices for the transaxial location of the photon (one for each crystal in the module, 2 virtual indices for the gaps on either side of the module), 55 indices for the axial location (9 crystals per detector block, and 2 virtual ones on either side of each block), and 5 degrees for both the transaxial and axial angles.
To take into account the fact that the MCGPU radius includes the DOI, the photon paths from the cylinder at the detector face were extended to a cylinder with the same radius as in MCGPU before binning the histogram.
Any photon that no longer fell within the axial extent of the scanner was discarded.

A similar 5D histogram was constructed based on the information from the singles in the GATE detectors.
Each of the photons was assigned to the same bin as in the actor histogram, based on its event ID.
Photons were added to the detector histogram if they were detected at all, the location of detection was not taken into account.
The detection probabilities in the LUT were calculated by dividing the detector histogram by the actor histogram.

To evaluate the effects of each LUT component, lower dimensional LUTs were computed by summing over the unused dimensions of the actor and detector histograms (taking into account whether coincidences in gaps would be kept during listmode processing or not), and then dividing them as before.
We created a 1D LUT using just energy information, one 3D LUT containing energy and angle information and another 3D LUT containing energy and crystal information.
The processing of coincidences hitting gaps in the MCGPU-PET listmode would be handled slightly differently for each LUT.
For LUTs without crystal information (or in the absence of a LUT), coincidences containing photons that hit a gap were discarded.
Therefore, photons hitting gaps were discarded from the actor and detector histograms before creating those LUTs.
For the 3D LUT with crystal information, photons hitting a gap would be assigned to the nearest edge crystal after receiving a probability.
For the 5D LUT, this assignment was done by a second LUT.

\subsubsection{Edge crystal assignment LUT}
\begin{figure}
    \centering
    \includegraphics[width=0.75\linewidth]{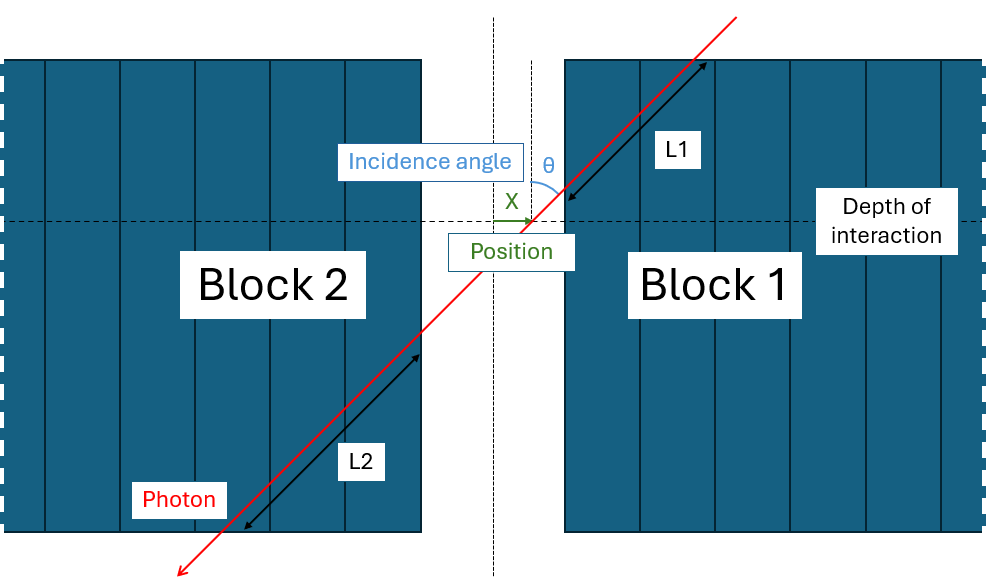}
    \caption{Schematic of two adjacent axial detector blocks in the scanner. The probability of a photon hitting a gap being detected in block 1 or block 2 is heavily dependent on the incidence angle of the photon, as well as the position at which it intersects the depth of interaction.}
    \label{fig:gap_assignment}
\end{figure}

The edge crystal assignment LUT was constructed to decide which edge crystal that photons intersecting the MCGPU-PET detector cylinder inside gaps would be assigned to.
The reason for this is shown in the schematic of two adjacent scanner blocks in Figure \ref{fig:gap_assignment}.
It is not trivial to see what the relative probability of detection is in either edge crystal when a photon hits the gap between two blocks. To calculate these probabilities for a given incidence angle and position, we computed the attenuation fraction using the attenuation coefficient of LYSO $\mu_\mathrm{LYSO}$ and the total distance $l_{n}$ that the photon spends inside either block, for a given incidence angle $\theta$ and position $x$.
For a positive $\theta$, the photon enters block 1 first, and the absolute ($p_{n,abs}$) and relative ($p_{n,rel}$) probabilities of detection are defined as follows:
\begin{equation}
    \label{eq:p1abs}
    p_{1,abs} = e^{-\mu_\mathrm{LYSO}\cdot l_{1}}
\end{equation}
\begin{equation}
    \label{eq:p2abs}
    p_{2,abs} = (1 - p_{1,abs}) \cdot e^{-\mu_\mathrm{LYSO}\cdot l_{2}}
\end{equation}
\begin{equation}
    p_{n,rel} = \frac{p_{n,abs}}{p_{1,abs} + p_{2,abs}}
\end{equation}
In the cases where $\theta$ is negative and the photon intersects block 2 first, it suffices to swap the block indices in equations \ref{eq:p1abs} and \ref{eq:p2abs}.
Two LUTs with probabilities were calculated, one for the axial and one for the transaxial gaps in the scanner, respectively.
For the transaxial gaps, the effect of the angle between the two modules was assumed to be negligible.
During listmode processing, after detection probabilities from the 5D LUT are assigned to each photon, photons intersecting the DOI inside a gap were assigned randomly to one of the edge crystals (not to any other crystals in the block), taking into account the relative probabilities of detection from the second LUT.
This ensures that on average, photons are assigned to the edge crystal of the correct block.
The reason for using an accept-reject policy instead of distributing the probabilities is because it would require adding copies of coincidences with different crystals to the list (even more so if both photons hit a gap), which would significantly complicate our implementation of the MCGPU-PET listmode processing.

In summary, the listmode processing (with the LUT) takes the original MCGPU-PET listmode (using cylinder coordinates) as the input, and outputs a new listmode file that uses crystals instead of coordinates.
This processing either assigns a weight to an event or discards it.
It never converts a single input event to more than one weighted output event, so the number of events never increases.
This is why gap events are not distributed over the edge crystals. During reconstruction or unlisting to a sinogram, the event contributes that weight, which is a value between 0 and 1.

\subsubsection{Scatter sensitivity sinogram}
The final scatter estimate sinograms were calculated by multiplying the output scatter sinogram from MCGPU-PET (which takes the respective LUT into account) with a scatter sensitivity sinogram.
This scatter sensitivity sinogram was calculated from another GATE simulation, using a 400 mm diameter cylinder with the same length as the scanner ($2.6 \cdot 10^{8}$ counts).
The scattered coincidences from this cylinder did not cover the entire sinogram, but the reduced diameter of the cylinder drastically reduced the real world simulation time needed to get an acceptable number of counts to compute the sensitivity correction sinogram with.
The cylinder was still larger than any of the three phantoms, and the reduced size did not cause any problems.
The scatter sinogram from this simulation was divided by the output scatter sinogram resulting from the simulation of the same phantom in MCGPU-PET, with the effects of the LUT applied.
Similar to the trues sensitivity sinogram, 14 radial symmetries were exploited to reduce the noise in both sinograms.
Since each LUT produces a different MCGPU-PET sinogram, four different scatter sensitivity sinograms were calculated, one for each LUT.
Since we assume that the object dependence is accounted for by the LUT, only a single scatter sensitivity sinogram per LUT is needed, as it should be valid for any object.

\subsection{Scatter estimation}
The performance of our approach for scatter estimation was evaluated in three scenarios of increasing difficulty. In all of these scenarios, we assume the attenuation image is known and correct.

\subsubsection{Known activity distribution, unlimited simulation time}
Initially, 5 approaches for scatter estimation were compared: no LUT, 1D (energy) LUT, 3D (energy, angles) LUT, 3D (energy, crystals) LUT and 5D (energy, angles, crystals) LUT.
The approaches were compared on scatter-compensated reconstructions for the all 3 phantoms (Figure \ref{fig:phantoms}).
The scatter sinograms for these reconstructions were generated from a known activity distribution, using very long MCGPU-PET simulations to generate a very high number of coincidences, so that the signal-to-noise ratio (SNR) would be as high as reasonably achievable.
The output sinograms from the simulator were created with the respective LUTs added to the simulator logic, and were multiplied with their respective scatter sensitivity sinograms to produce the final scatter sinogram for each approach (Figure \ref{fig:schematic_sensitivities}).
The evaluation metrics and reconstruction parameters are detailed below.

\subsubsection{Known activity distribution, very limited simulation time}
We selected the 5D LUT for use in the second scenario, due its superior performance over the other LUTs.
We then investigated the effect of low-count scatter estimates on the reconstruction accuracy, to see how much we could reduce the simulator runtime.
We found that noise in the scatter estimates leads to bias in the reconstructions, meaning noise reduction would be required to be able to use low-count scatter estimates.
This observation is consistent with the results of other investigations into the effect of noise in the additive contribution sinogram on the reconstruction \cite{Hogg2003,Beekman1997,King1997}.
The cause of the effect is not known, but may be related to the non-negativity constraint of MLEM, as NEGML reconstructions are less biased for the same noise level in the randoms sinogram \cite{VanSlambrouck2015}.
For noise reduction, Gaussian smoothing was preferable over increasing simulation time, as it is much cheaper in terms of the runtime needed for a similar SNR increase.
However, the scatter estimates contain many high frequency structures as a result of the detector efficiency modeling using the 5D LUT, which complicates smoothing.

To get around this, we defined a single ``smoothing normalization sinogram'' for all phantoms.
The output sinogram from the simulator was multiplied with the smoothing normalization sinogram to remove most high frequency structures before smoothing.
The structures were re-applied after smoothing by dividing by the same normalization sinogram.
This smoothing normalization sinogram was calculated by taking a heavily smoothed version of the MCGPU-PET scatter sinogram for the big 40 cm cylinder and dividing it by the original sinogram.
After smoothing, the output sinogram was multiplied with the scatter sensitivity sinogram to produce the scatter estimate.

We tested different combinations of simulated count levels and FWHMs for smoothing.
The reconstructions made using the resulting scatter estimates were then compared to reconstructions using the GATE trues-only sinogram and the high count scatter sinogram from scenario 1 (Figure \ref{fig:smoothing_recons}).
We selected the combination of count level and smoothing amount (with the lowest runtime) that could still achieve a global bias $<1\%$ and a local bias $<3\%$ in the reconstructions for all phantoms, when compared to the scenario 1 reconstructions.
We also required the smoothing FWHM to be equal to or smaller than that of the highest frequencies we observed in the scatter sinograms of the three phantoms.
We also determined the highest frequencies present in the scatter sinograms using Richardson-Lucy deconvolution, which served as a soft upper limit for the smoothing FWHMs. The kernel widths we obtained were 4 cm, 8° and 5 cm, or 10 radial, 5 angular and 18 planar sinogram bins, respectively.

\subsubsection{Unknown activity distribution, very limited simulation time}
The third scenario required joint estimation of the activity distribution and the scatter sinogram, which we performed iteratively.
The iterative scatter estimation pipeline consisted of the following steps: (1) reconstruct activity image with current scatter estimate, (2) run short MCGPU-PET simulation with current activity and known attenuation image, (3) reduce noise in new scatter estimate, (4) return to (1) and repeat until desired number of iterations is reached.
For steps 2 and 3, we used the same low-count simulation and smoothing approach as in scenario 2. 

10 iterations of scatter estimation were run for all three phantoms, during which reconstructions were performed the same way as for the simulations with known activity input, but only for 1 OSEM iteration and 28 subsets.
An initial (heavily biased) input activity image was obtained from a reconstruction of the acquisition sinogram (sinogram from GATE containing trues and scatters) without scatter compensation.
The attenuation image was the voxelized representation of the GATE attenuation image (which was also used to compute the attenuation sinogram).

\subsubsection{Evaluation metrics}
The accuracy of the estimated scatter sinograms in the first scenario was compared by computing difference images for the mean over all direct sinogram planes (ring difference of 0), using the GATE scatter sinogram as a reference.
Relative differences were computed as
\begin{equation}
    (A-B)/B
\end{equation}
with A being the image to be compared and B the reference image. For pixels where B was 0, the relative difference was set to 0.
For absolute differences, the division by the reference image was omitted.

In all scenarios, we performed scatter-compensated reconstructions using TOF-OSEM \cite{Hudson1994} with resolution modeling, using 4 iterations and 28 subsets. The reconstructed image dimensions were $192 \times 192 \times 90$ with voxel sizes of $2.734 \times 2.734 \times 2.890\ \mathrm{mm^{3}}$. The attenuation sinogram was derived from a forward projection of a voxelized representation of the attenuation map used in the GATE simulation.

\begin{figure}
    \centering
    \includegraphics[width=0.75\linewidth]{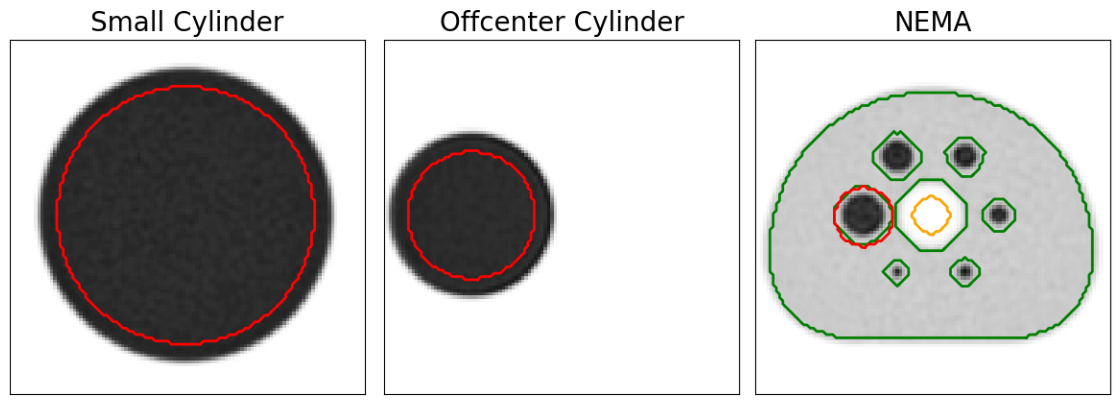}
    \caption{Visualization of ROIs on trues only reconstructions. Average over all planes for the cylinders, single plane for the NEMA phantom. Green indicates the large background ROI, red is the ROI for the biggest sphere, and orange is the cavity ROI.}
    \label{fig:rois}
\end{figure}

Reconstruction accuracy was evaluated by comparing to a reconstruction made using only true coincidences (Trues only, TO), to the ground truth image (GT), or to the reconstruction made with the high count scatter estimate (HC) from scenario 1.
Relative difference images were made between the reconstructions, by taking the mean over all slices, subtracting the reference image and dividing by the ground truth phantom concentration (background concentration for the NEMA phantom).
A quantitative comparison was made using regions-of-interest (ROIs) for all three phantoms (Figure \ref{fig:rois}).
For the cylindrical phantoms, a cylindrical ROI was defined with the same length as the phantom, and a radius smaller than the real radius by $10\times$ the reconstruction resolution, to avoid the Gibbs artifacts at the edges.
For the NEMA phantom, three ROIs were defined: a spherical region centered on the largest sphere, a cylindrical region for the air cavity, and a large background region.
In these ROIs, we calculated MSE as the mean of the squared voxel-wise difference (no averaging over slices) with the reference reconstruction.
The global bias was calculated as
    $$(C-D)/E$$
with C the mean ROI value of the reconstruction to be compared, D the mean ROI value of the reference reconstruction, and E the phantom concentration (background concentration for the NEMA phantom).
Local biases were computed by dividing the reconstruction into $12\times12\times12$ cubic ROIs and computing the bias using the same formula as for the big ROIs.
Any cubes with an average activity concentration lower than 500 Bq/ml were not considered during evaluation.

\subsection{Scatter estimation for phantom scans on the GE Signa PET/MR}
After the quantitative evaluation on the simulated acquisitions, we also applied our scatter estimation approach to real acquisitions, as a proof-of-concept. For this purpose, we acquired two scans for phantoms similar to the ones in the simulated evaluation. One big difference between these acquisitions and the simulated ones was the presence of random coincidences, for which a separate expectation sinogram was estimated using vendor-provided software. The scanner geometry, energy resolution, energy window and TOF resolution were identical to the simulated acquisitions.

\subsubsection{Small cylinder acquisition}
The first phantom acquisition consisted of a small cylinder with a length of 20 cm and a radius of 10 cm, placed in the center of the scanner's FOV. It was filled with 0.9 kBq/ml of F-18 at the start of the scan. Total scan time was 2 hours, resulting in $3.74 \cdot 10^8$ prompt coincidences.

\subsubsection{NEMA phantom acquisition}
The second phantom acquisition consisted of a NEMA image quality phantom, placed in the center of the scanner's FOV. The smallest and largest spheres were filled with an activity ratio of 4:1 compared to the background. The exact F-18 activity concentration was unfortunately not known, but estimated to be approximately 2.5 kBq/ml for the background and 10 kBq/ml for the spheres. Total scan time was 1 hour, resulting in $1.22 \cdot 10^9$ prompt coincidences.

\subsubsection{Scatter estimation for real acquisitions}
To estimate scatter for real acquisitions, we had to account for several effects/phenomena present in our scanner, but not yet modeled during our simulated evaluation:
\begin{enumerate}
    \item Decay: The effect of activity decay was not directly applied during the MCGPU-PET simulation, but rather incorporated as a single scale factor based on the length of the scan, for practical reasons.
    \item Random coincidences: The randoms expectation sinogram was estimated using vendor software, and included in the forward model during reconstruction.
    \item Inter-crystal scatter recovery: In the real scanner, photons that scatter inside the detector crystals between neighboring blocks are recovered \cite{Hsu2019,Wagadarikar2012}, an effect we did not simulate in our GATE acquisitions. The resulting sensitivity increase for the scanner was incorporated in the LUT by including the recovered photons (in GATE) in the probability calculations, and the corresponding scatter sensitivity sinogram was obtained by re-running the required simulations with the updated LUT and GATE logic.
    \item Dead-time and pile-up: The vendor estimates for dead-time and pile-up were included in the simulator's forward model.
    \item Different relative detector sensitivities: The vendor estimates for non-uniformity of the detector sensitivities were included in the simulator's forward model.
    \item Correction factor for non-ideal detectors: In our testing, even with all the modifications above, the ratio between the expected number of scattered coincidences for both phantom acquisitions and the number we simulated was only around \%75. We assume this is due to the fact that the detectors in GATE (which also determine the detection probabilities in the LUT) are an ideal version of those in the scanner, and that this difference should be captured by the trues sensitivity sinogram of the scanner. Since the sensitivity of the scanner is not constant over time, we estimated one global scale factor per acquisition from the ratio of the GATE trues sensitivity sinogram and the (acquisition-specific) scanner trues sensitivity sinogram, which assumes that this global scale factor is the same for true and scattered coincidences. The GATE trues sensitivity sinogram used for this purpose was different from the one used for reconstructing our simulated acquisitions, as it includes corrections for the effects mentioned above (similar to the scanner trues sensitivity sinogram). The resulting correction factors were 0.786 for the cylinder and 0.730 for the NEMA phantom, respectively.
\end{enumerate}
Apart from these modifications, the scatter simulation and joint estimation approach was identical to the one used in the third scenario of the simulated evaluation (unkown activity, short simulation time).

\subsubsection{Comparison to tail-fitted SSS}
Aiming for the fairest possible comparison between tail-fitted SSS and MC simulated scatter, we kept as much as possible of the reconstruction pipeline identical between both approaches, and only replaced the scatter estimate. The randoms estimate, the trues sensitivity sinogram, the attenuation sinogram etc. were the same for both approaches. The reconstruction parameters were identical to those in the simulated evaluation (i.e. TOF-OSEM, 4 iterations and 28 subsets, $192 \times 192 \times 90$ voxels of $2.734 \times 2.734 \times 2.890\ \mathrm{mm^{3}}$). Since the Signa is a PET/MR scanner, attenuation maps for both phantoms had to be obtained from CT-derived attenuation templates, and had to be combined with an attenuation template of the scanner bed. The estimated scatter sinograms for both approaches were compared by plotting non-TOF radial profiles.

%% file: results.tex
\section{Results}

\subsection{Known activity distribution, unlimited simulation time}
\begin{figure}
    \centering
    \includegraphics[width=0.75\linewidth]{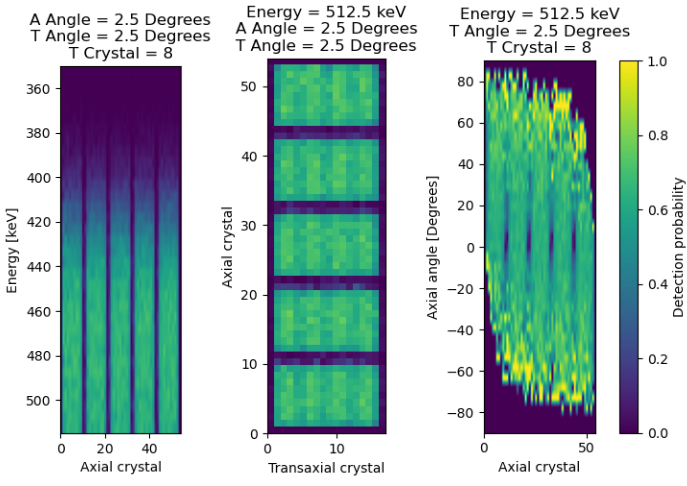}
    \caption{2D slices through the 5D LUT, which assigns probabilities of detection to single photons based on their energy, axial and transaxial incidence angle, and axial and transaxial crystal index. Centers of the bins which were kept fixed for each slice are shown at the top. Note: some crystals are virtual, meant for photons detected between detector blocks.}
    \label{fig:lut}
\end{figure}

\begin{figure}
    \centering
    \includegraphics[width=1\linewidth]{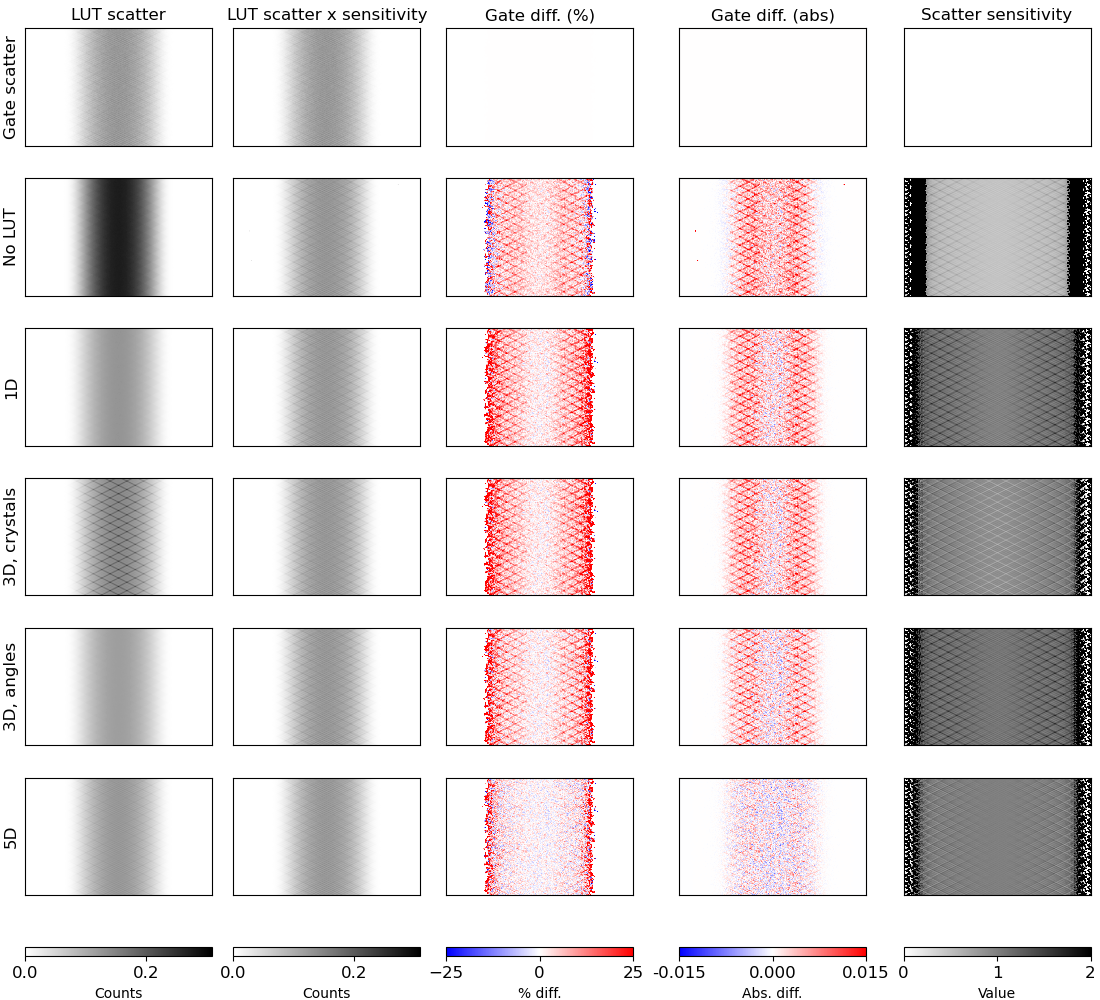}
    \caption{Mean over direct planes for small cylinder sinograms directly after MCGPU listmode processing with the LUT (LUT scatter), and after multiplying with the respective scatter sensitivity sinogram (LUT scatter $\times$ sensitivity). The third and fourth columns show a comparison between the scatter estimates and the GATE scatter sinogram. The last column shows the corresponding scatter sensitivity sinograms for each LUT.}
    \label{fig:sinograms}
\end{figure}

In the first scenario, we tested the combination of different LUTs with their respective scatter sensitivity sinograms for scatter estimation when the activity distribution is known, and when it is possible to use long, high count MC simulations to obtain high-SNR scatter estimates.
Figure~\ref{fig:lut} shows cross-sections through the 5D detection probability LUT.
The MCGPU-PET scatter simulations produced $6 \cdot 10^{10}$ scatter coincidences in each simulation, which translates to around 36 hours of real-world simulation time per phantom and per LUT on our Nvidia A6000 GPUs.
The intermediate and final scatter sinograms for the small cylinder are shown in Fig.~\ref{fig:sinograms}.
This figure also shows a comparison with the scatter sinogram from the GATE simulation, and the respective scatter sensitivity sinograms for each LUT.
The scatter estimate generated using the 5D LUT and scatter sensitivity shows the smallest differences with the GATE scatter sinogram.

\begin{figure}
    \centering
    \includegraphics[width=1.0\linewidth]{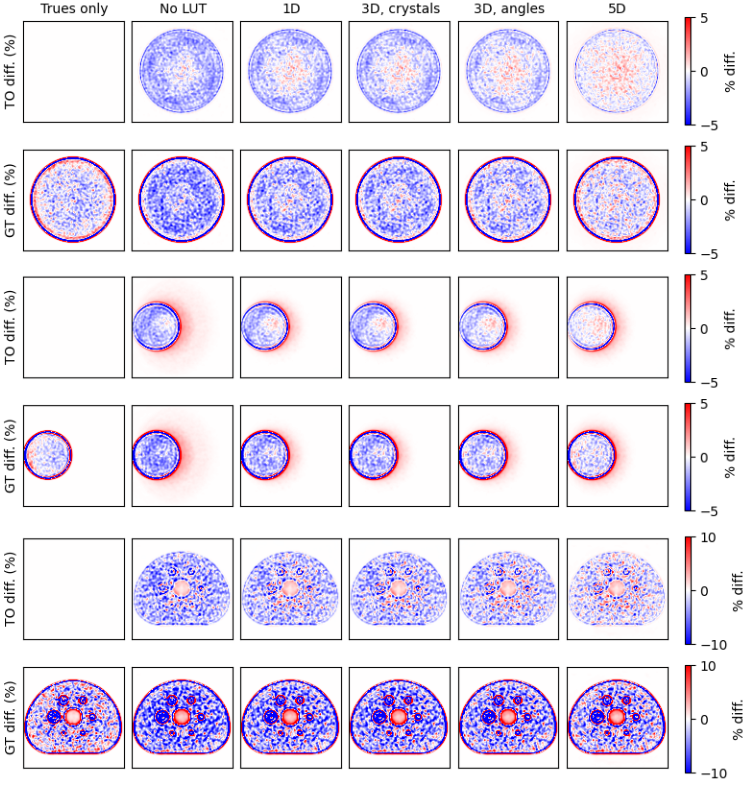}
    \caption{Differences between reconstructions of all phantoms made using MC-simulated scatter sinograms (with different LUTs and scatter sensitivities), the trues only reconstruction (TO) and the ground truth image (GT). The differences are expressed as a percentage of the phantom concentration (background concentration for the NEMA phantom). Recons are averaged over all planes for the cylindrical phantoms, a single slice is shown for the NEMA phantom. The differences are expressed as a percentage of the phantom concentration. Note the different scales.}
    \label{fig:recons}
\end{figure}

Scatter-compensated reconstructions were performed using the scatter estimates for each LUT and compared to the trues only reconstruction and the ground truth activity image (Figure~\ref{fig:recons}).
The accuracy of the scatter-compensated reconstructions was also evaluated quantitatively, by computing the bias and MSE within the defined ROIs (Figure~\ref{fig:rois}) with the trues-only reconstruction as the reference. These ROI values are summarized in Table~\ref{tab:roi_values}.

\begin{table}
\begin{tabular}{lllllllllll}
    \toprule
     & \multicolumn{2}{c}{Small Cylinder} & \multicolumn{2}{c}{Offcenter Cylinder} & \multicolumn{6}{c}{NEMA} \\
    \midrule
    Metric & MSE & Bias (\%) & MSE & Bias (\%) & \multicolumn{3}{c}{MSE} & \multicolumn{3}{c}{Bias (\%)} \\
    Region & CL & CL & CL & CL & BS & BG & CA & BS & BG & CA \\
    No LUT & 3.054 & -1.20 & 2.105 & -1.18 & 1.543 & 0.644 & \textbf{0.074} & -3.02 & -2.09 & 1.20 \\
    1D & 2.956 & -0.69 & 1.992 & -0.74 & 1.252 & 0.549 & 0.076 & -1.71 & -1.54 & 1.21 \\
    3D, crystals & 2.946 & -0.68 & 1.980 & -0.70 & 1.245 & 0.549 & 0.076 & -1.73 & -1.55 & 1.20 \\
    3D, angles & 2.906 & -0.41 & 1.910 & -0.44 & 1.197 & 0.509 & 0.076 & -1.36 & -1.24 & 1.20 \\
    5D & \textbf{2.826} & \textbf{0.12} & \textbf{1.812} & \textbf{0.06} & \textbf{1.144} &\textbf{ 0.448} & 0.077 & \textbf{-0.85} & \textbf{-0.63} & 1.21 \\
    \bottomrule
\end{tabular}
    \caption{MSE and bias of the activity values compared to the trues only reconstruction per ROI, for each phantom and LUT combination. Values in bold indicate the lowest MSE and smallest bias for each region. (CL: Cylinder, BS: Biggest sphere, BG: Background, CA: Cavity)}
    \label{tab:roi_values}
\end{table}

\subsection{Known activity distribution, very limited simulation time}
\begin{figure}
    \centering
    \includegraphics[width=\linewidth]{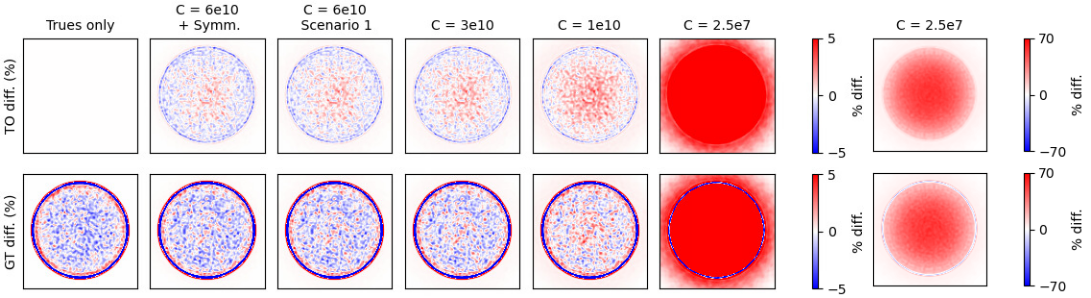}
    \caption{Differences between recons of the small cylinder phantom made using MC-simulated scatter sinograms (with known activity and different count levels (C)), the trues only reconstruction (TO) and the ground truth image (GT). The differences are expressed as a percentage of the phantom concentration. Recons are averaged over all planes. The SNR was also increased beyond that of the 6e10 counts sinogram by exploiting 14 radial symmetries (+ Symm). The 2.5e7 counts reconstruction is shown again separately, using a more fitting color scale.}
    \label{fig:snr_recons}
\end{figure}

For the second scenario, we first investigated the effect of the number of MC simulated counts on the scatter estimates by performing reconstructions with scatter estimates of different count levels.
The scatter-compensated reconstructions in Fig.~\ref{fig:snr_recons} show that lower count levels (lower SNR) in the MCGPU-PET simulations increases the bias in the resulting reconstructions.

\begin{figure}
    \centering
    \includegraphics[width=0.75\linewidth]{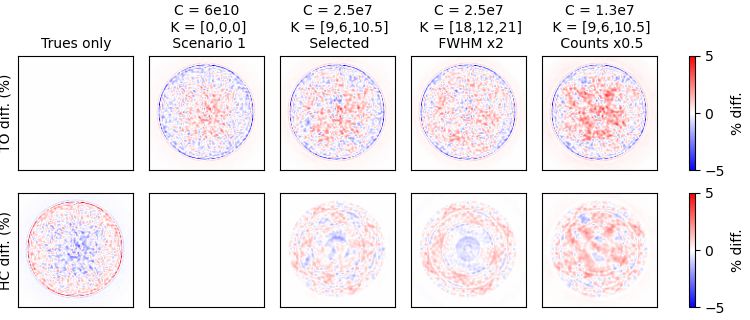}
    \caption{Differences between recons of the small cylinder phantom made using MC-simulated scatter sinograms (with known activity and different combinations of count level (C) and smoothing intensity (kernel width, K)), the trues only reconstruction (TO) and the reconstruction from scenario 1 (HC). The differences are expressed as a percentage of the phantom concentration.}
    \label{fig:smoothing_recons}
\end{figure}

The SNR in the scatter estimates was improved through Gaussian smoothing.
The smoothing was implemented on the same GPUs as the MCGPU-PET simulation, and took around 2 min. per sinogram, independent of FWHM.
Multiple combinations of count levels and smoothing FWHMs were tested against our predefined criteria (global bias $<1\%$ and local bias $<3\%$ compared to the reconstruction from scenario 1).
The lowest count level that could still achieve this was $\sim2.5\cdot10^{7}$, which took around 40 s of real world time to simulate (not including listmode processing time). The subsequent Gaussian smoothing was performed using a FWHM of $[9,6,10.5]$ radial, angular and planar bins.
Figure \ref{fig:smoothing_recons} shows the increases in local and global bias in the resulting reconstructions as compared to scenario 1 when higher values for the FWHM or lower numbers of simulated counts were chosen.

\subsection{Unknown activity distribution, very limited simulation time}
\begin{figure}
    \centering
    \includegraphics[width=0.75\linewidth]{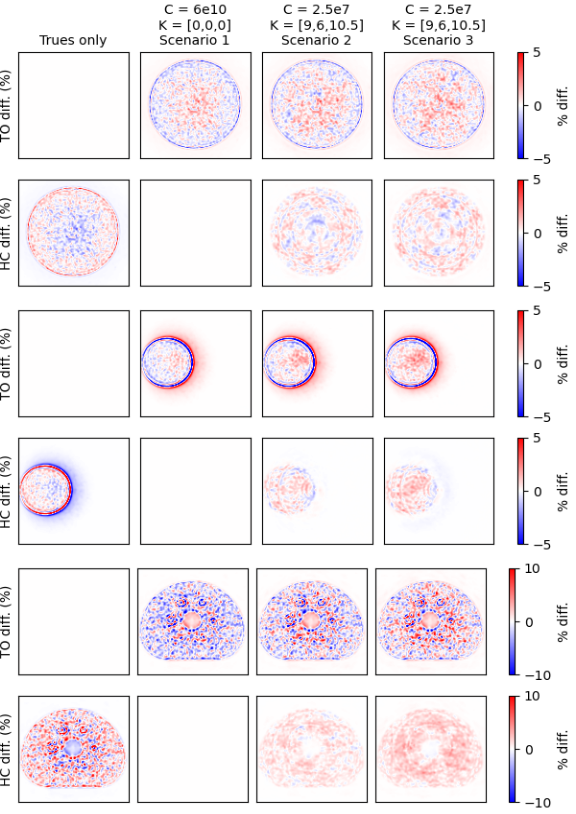}
    \caption{Differences between recons made using MC-simulated scatter sinograms from all 3 scenarios, the trues only reconstruction (TO) and the reconstruction from scenario 1 (HC). The differences are expressed as a percentage of the phantom concentration. (C = Counts in the simulation, K = Smoothing kernel width)}
    \label{fig:iterative_recons}
\end{figure}

The iterative scatter estimation in scenario 3 was run for 10 iterations, using the number of simulated counts and the FWHMs selected in scenario 2.
The resulting scatter estimate was used to create the scenario 3 reconstruction.
Figure \ref{fig:iterative_recons} shows a comparison between the trues only reconstruction (TO) and the reconstructions from scenarios 1 (HC), 2, and 3.
The scenario 3 MSE and bias for each ROI and phantom were calculated using the scenario 1 reconstruction as the reference (Table \ref{tab:iter_values}).

\begin{table}
\begin{tabular}{lllllllllll}
    \toprule
     & \multicolumn{2}{c}{Small Cylinder} & \multicolumn{2}{c}{Offcenter Cylinder} & \multicolumn{6}{c}{NEMA} \\
    \midrule
    Metric & MSE & Bias (\%) & MSE & Bias (\%) & \multicolumn{3}{c}{MSE} & \multicolumn{3}{c}{Bias (\%)} \\
    Region & CL & CL & CL & CL & BS & BG & CA & BS & BG & CA \\
    Trues only & 2.826 & -0.12 & 1.812 & -0.06 & 1.144 & 0.448 & 0.077 & 0.85 & 0.63 & -1.21 \\
    Scenario 2 & 0.513 & 0.13 & 0.483 & 0.34 & 0.075 & 0.033 & 0.002 & 0.96 & 0.31 & -0.08 \\
    Scenario 3 & 0.487 & 0.22 & 0.465 & 0.63 & 0.234 & 0.064 & 0.001 & 2.13 & 0.82 & 0.04 \\
    \bottomrule
\end{tabular}
    \caption{MSE and global bias of the activity values in the trues only, scenario 2 and scenario 3 recons, compared to the scenario 1 reconstruction for each phantom and per ROI. (CL: Cylinder, BS: Biggest sphere, BG: Background, CA: Cavity)}
    \label{tab:iter_values}
\end{table}

\subsection{Real scans on the Signa PET/MR}
\begin{figure*}[t!]
    \centering
    \begin{subfigure}[t]{\textwidth}
        \centering
        \includegraphics[width=\linewidth]{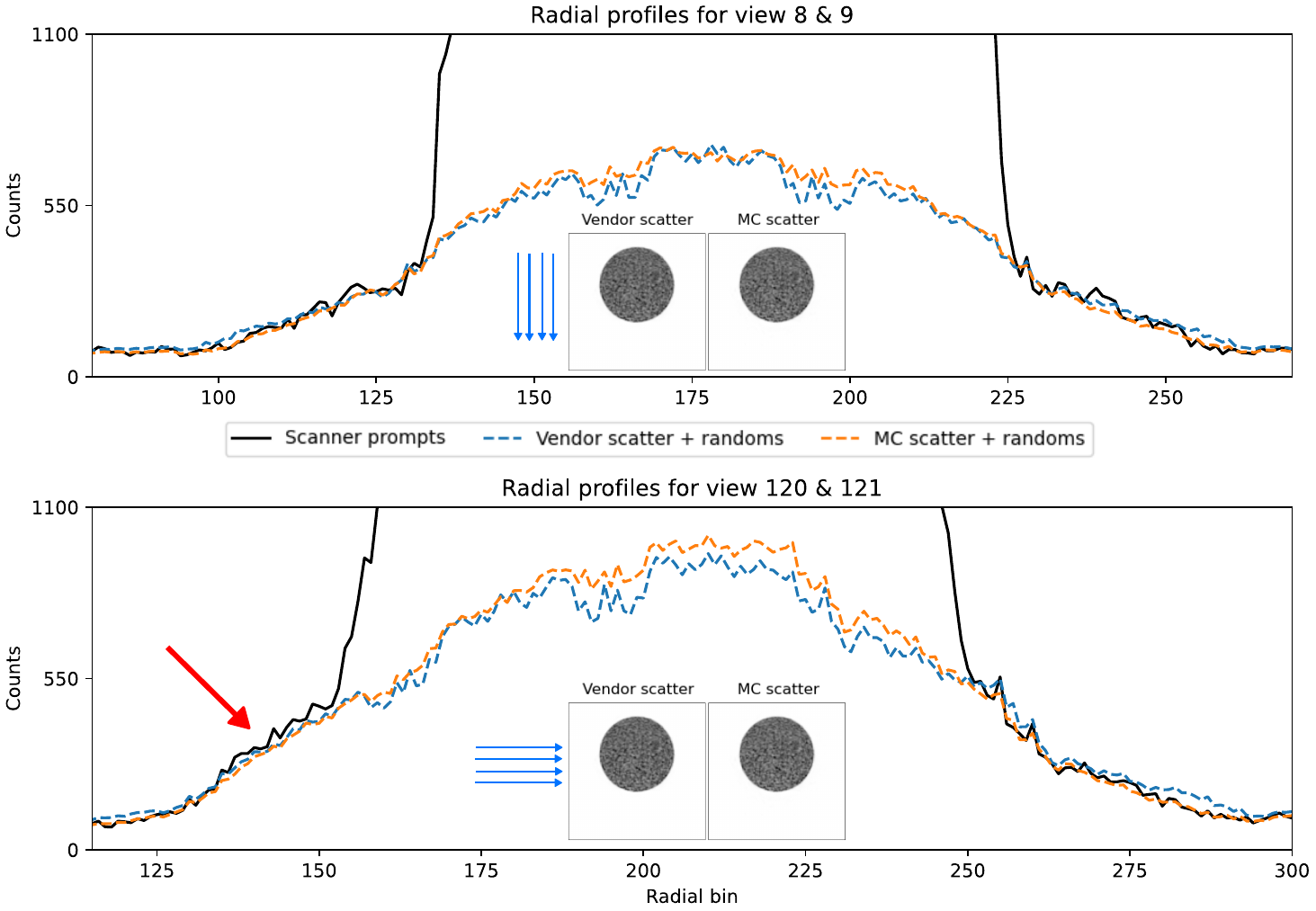}
        \caption{Cylindrical phantom.}
        \label{fig:cylinder_real}
    \end{subfigure}%
    
    \begin{subfigure}[t]{\textwidth}
        \centering
        \includegraphics[width=\linewidth]{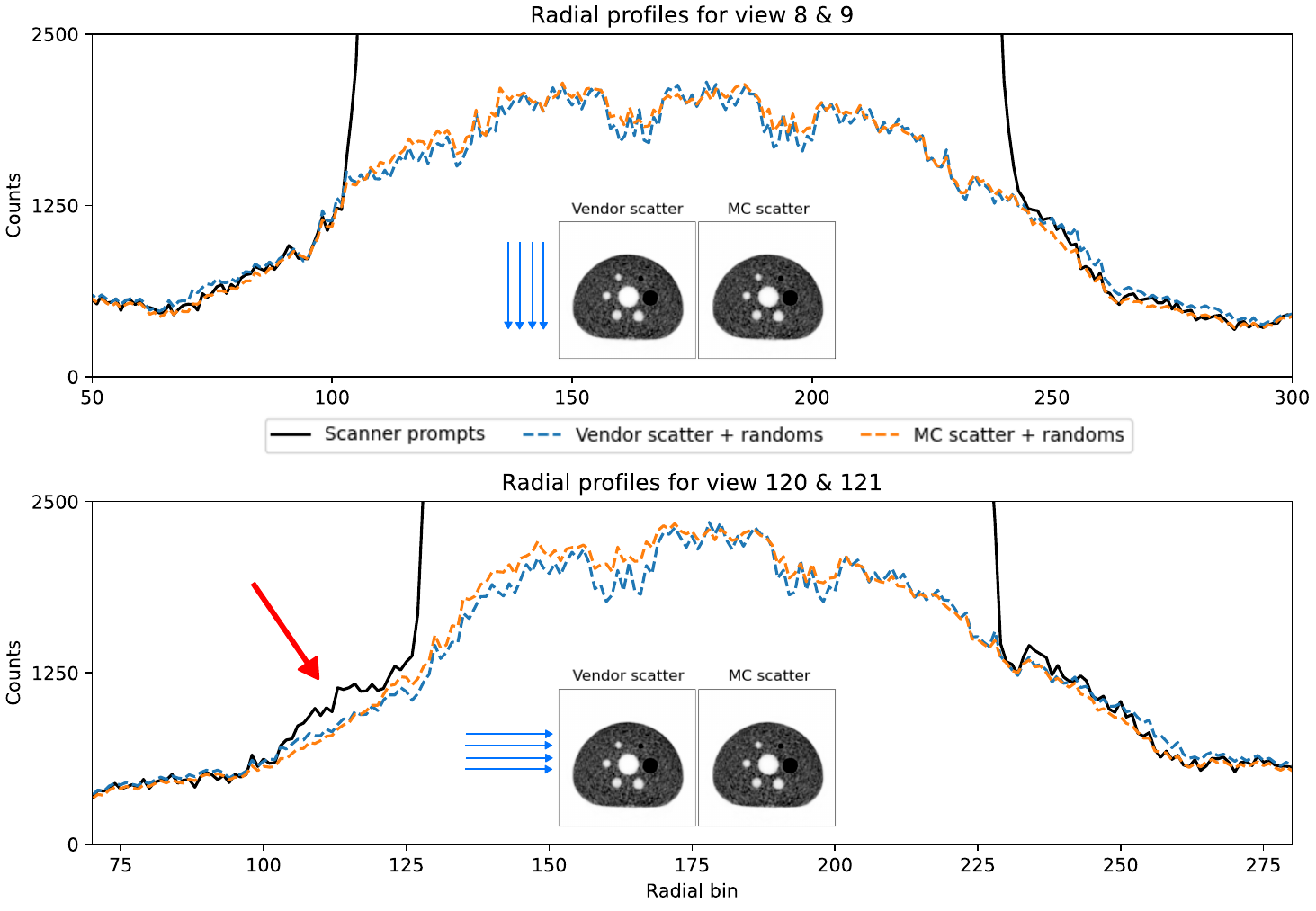}
        \caption{NEMA phantom.}
        \label{fig:nema_real}
    \end{subfigure}
    \caption{Plots of radial profiles through non-TOF sinograms averaged over all planes and 2 views, for the prompts and contamination sinograms. Large red arrows indicate mismatches between the prompts profile tail and both contamination profiles. Reconstructions made using the vendor scatter (SSS) and MC-simulated scatter are shown at the bottom. Small blue arrows show the orientation of the LORs contained in the respective views.}
    \label{fig:real_acquisitions}
\end{figure*}

The iterative scatter estimation for the real scans was run for 10 iterations, using the same parameters for smoothing and the number of simulated counts as scenarios 2 and 3 in the simulated evaluation. The scatter estimation, including the final reconstruction, took around 3 hours. Of those 3 hours, 5 minutes were spent on MC simulation and 20 minutes on smoothing of the scatter sinograms. Reconstructions using the vendor and MC-simulated scatter, as well as sinogram profiles for the prompts (true, scattered and random coincidences) and contamination (scattered and random coincidences) sinograms are shown in Fig.~\ref{fig:real_acquisitions}. There are no apparent differences between the MC and vendor scatter reconstructions, and both contamination sinogram profiles also agree very well in most radial bins.

%% file: discussion.tex
\section{Discussion}
\subsection{Detector efficiency LUT}
The 5D LUT obtained from the GATE simulations was still slightly noisy, though this did not have a noticeable effect on the sinograms it produced. The LUT also captured some interesting effects, which can be seen in the 2D slices in Fig.~\ref{fig:lut}:
\begin{itemize}
    \item Incident photon energies down to about 400 keV still have a 10\% chance of being detected, because of the imperfect energy resolution of the detectors.
    \item While not really visible in the figure, the detection probability slopes slightly downwards for photon energies higher than 475 keV.
    \item Probability of detection is slightly lower for photons hitting crystals on the edges of detector blocks instead of the center, due to photons undergoing inter-crystal scatter being irrecoverable outside the detector block.
    \item Photons hitting the detectors at an oblique angle have a slightly higher chance of being detected than photons hitting the detector straight on, because of the increased path length inside the detectors.
    \item When photons hit between blocks, their probability of detection is even more dependent on their incidence angle, as the gap between the detectors causes it to have an even bigger impact on how long the photon path inside the detectors is.
\end{itemize}

\subsection{Known activity distribution, unlimited simulation time}
We found that the combination of the 5D LUT and a scatter sensitivity sinogram produces the closest scatter estimate to the GATE scatter sinogram, even when the scatter sensitivity sinogram is derived from a different phantom.

The shortcomings of the other LUTs seem to be the LORs connecting to crystals at the edges of the detector blocks, which can be attributed to the coincidences where at least one photon hits a gap between detector blocks.
In this scanner geometry, these photons account for about 20\% of all simulated coincidences, which is a sizable portion.

One contributing component to the shortcomings of the lower dimensional LUTs could be the way the photons are assigned to the edge crystals (or discarded), but a more likely explanation is that the detection probability for these photons is highly dependent on their incidence angle, the distribution of which is highly dependent on object size.
If not correctly scaled by the LUT, coincidences containing these photons will bias the value in the sensitivity correction sinogram. 
The information about photon hit location and photon angle of incidence (necessary to identify these photons and assign the correct detection probability) are only present together in the 5D LUT, which explains why it manages to correct for it.

When applied to reconstructions, the 5D LUT consistently had the lowest bias and MSE of all the LUTs compared to the trues only reconstruction, and had the fewest artifacts.
While not easily visible, all recons except the one from the 5D LUT contained ringing artifacts, which are likely related either to the issue with LORs connecting edge crystals explained above, or to the fact that the angular FWHM (6 bins) is slightly higher than our upper limit (5 bins).

The comparison with the GT is only shown in Fig.~\ref{fig:recons} and \ref{fig:snr_recons} and left out of the ROI comparison, as we believe it is of little added value.
The reason is that our trues only reconstruction contains a slight negative bias $<0.5\%$ compared to the GT activity image, likely due to an attenuation modeling mismatch between our forward model and the GATE simulation.
We found a similar bias when performing a scatter-compensated reconstruction both using the exact GATE scatter sinogram and when using a higher SNR scatter sinogram with 14 radial symmetries applied. 
Therefore, we believe it is more logical to compare the scenario 1 reconstructions to the trues only reconstruction, as the other reconstructions will contain a similar bias caused by this forward model mismatch.

The leftover positive bias in the 5D LUT reconstruction has two possible causes related to the scatter sensitivity sinogram.
One possible explanation is the fact that the scatter sensitivity sinogram is also slightly noisy, because it is calculated from a GATE scatter sinogram. This noise then propagates into the scatter estimates, and as demonstrated in Fig.~\ref{fig:snr_recons}, even very small amounts of noise can cause bias in the reconstructions. 
However, running the simulation that produces the GATE scatter sinogram for longer to get a better SNR was not feasible, as the runtime for such high-SNR GATE simulations quickly reaches weeks or months.
Another possibility is that the LUT is missing an object-dependent effect which in turn creates bias in the scatter sensitivity sinogram. 

Indeed, it is not realistic to expect that a LUT in combination with a sensitivity correction sinogram can perfectly model the physics that are simulated in advanced MC simulators like GATE or SimSet.
However, our method for incorporating the scatter sensitivities is a valid approximation for simulators which model a simple cylindrical scanner and do not simulate detector physics (i.e. detect all events hitting the cylinder).
It can even be extended to different scanner geometries, which would only require repeating the simulations for the LUT and trues and scatter sensitivity sinograms with the new geometry.
While our approach can provide a significant speedup of MC simulators by skipping the simulation of photon-detector interactions, the value of a fast simulator which \textit{can} simulate detector physics should not be underestimated. We believe that once the speed of simulators with detector physics reaches a certain point, their superior accuracy will be preferable over the shorter simulation time of the detection probability LUT and sensitivity sinogram approach.

\subsection{Known activity distribution, very limited simulation time}
The reconstructions from scenario 2 show that it is possible to reach similar scatter compensation performance to a high count MC simulated scatter estimate (scenario 1) with a low count scatter estimate, provided that the noise in the sinogram is reduced first (e.g. using Gaussian smoothing).

Figure~\ref{fig:smoothing_recons} shows that increasing the smoothing beyond the selected values introduces artifacts which increase local bias, and that reducing the number of counts beyond the selected value increases both local and global bias beyond our selection criteria.

The scenario 1 reconstruction was chosen as the reference, because even that reconstruction already contains some bias, and we wanted to avoid that bias caused by the smoothing approach would compensate the already present bias, and mistakenly show really good performance when compared to the trues only reconstruction.

\subsection{Unknown activity distribution, very limited simulation time}
For this proof-of-concept, we purposefully kept the implementation of our iterative scatter estimation pipeline in scenario 3 simple, as the implementation was not the main focus of this paper.
This means there are some aspects of the pipeline which can still be improved:
\begin{itemize}
    \item Of the 40s spent simulating scatter in each iteration, about half was spent in initialization, which (among other factors) depends mostly on the size of the input images.
    \item The number of simulated counts could be reduced further if the reconstructions during iterative scatter estimation were performed using bigger pixels, which reduces noise (and reconstruction time as well).
    \item Mashing the sinograms (combining LORs) would further reduce the minimum required number of counts, as well as the amount of time needed for smoothing.
    \item Currently, the sinograms are smoothed separately in each segment (set of planes), which drastically reduces the advantage of GPU-based smoothing.
    If the sinograms could be structured in such a way that they can be smoothed entirely in a single pass, this would significantly reduce smoothing time.
\end{itemize}

The smoothing approach and the included normalization were purposefully kept simple as well, for the same reasons as the iterative scatter estimation. The ``smoothing normalization sinogram'' is not entirely object independent, and leaves some high-frequent structures in the sinograms after multiplication. Smoothing them away causes reconstruction artifacts to appear, as demonstrated in Fig.~\ref{fig:smoothing_recons}. More sophisticated normalization and/or smoothing methods may be able to avoid smoothing away these structures.

Nevertheless, the results of scenario 3 remained within 1\% global bias and 3\% local bias of the scenario 1 reconstruction (the same criteria as for scenario 2), with the exception of the biggest sphere of the NEMA phantom.

\subsection{Real scans on the Signa PET/MR}
Based on the reconstruction slices shown in Fig.~\ref{fig:real_acquisitions}, it is very hard to tell the difference between the reconstruction made using the vendor scatter, and the one made using MC-simulated scatter. In the sinogram profiles, we can see that the MC-based scatter sinogram follows the tails of the prompts sinogram (where only scatter is expected) very well. However, for the bins indicated with the red arrows, both contamination profiles fail to follow the prompt tails. These bins correspond to LORs passing through the bed of the scanner, so the mismatch might be explained by an imperfect estimate of the attenuation, which is plausible when using template-based attenuation images. At certain positions in the tails, the MC scatter performs better than the vendor scatter, which sometimes slightly overestimates the number of scatters. However, while the intrinsic scaling of the MC-based scatter estimation gives it a definitive advantage over tail-fitted SSS, it is impossible to make any statements about which is better (and under which conditions) without a proper reference and investigation. The results presented here merely serve to prove that it is possible to estimate scatter for real scans using MC simulation in a reasonably short time, and that the resulting sinograms and reconstructions are close to those produced by tail-fitted SSS (i.e. the current standard for clinical practice).

\subsection{Future work}
The scatter estimation pipeline we have presented here is a proof-of-concept, and there are a few aspects that future work could improve upon. One important point is the speed of the iterative scatter estimation approach and the processing of the MC simulation outputs, which we purposefully did not optimize, as we only aimed to show that scatter estimation can still be done accurately with fast, low-count MC simulations. Another point of improvement is the noise reduction step in the pipeline, as the current approach is fairly simple and may introduce additional artifacts. For real scans, it would also be useful to further investigate the reason for the differences between the GATE and scanner trues sensitivities, and to determine whether a single global scale factor is a reasonable approximation.

While out of the scope of this work, a deeper dive into the accuracy, robustness and speed of MC-based scatter estimation (as compared to tail-fitted SSS) will be essential if MC-based scatter estimation is ever to become the new clinical standard.

%% file: conclusion.tex
\section{Conclusion}
We showed that when estimating scatter using a simple cylindrical simulator without detector physics, a scatter sensitivity sinogram estimated for a given object should not be used for other objects, especially those of different sizes.
This object dependence can be reduced through the addition of a sensitivity at the level of single photons by incorporating a 5D LUT in the simulator logic, which assigns detection probabilities to individual photons depending on photon energy, incidence angle and crystal location.
Essentially, the 5D LUT provides a simple way to extend the MC simulator with a crude model of the detector physics.

We applied this approach to the MCGPU-PET MC simulator (which does not simulate detector physics), and used it to estimate the scatter for a virtual GATE-simulated acquisition in three scenarios of increasing difficulty:
\begin{enumerate}
    \item In the presence of a known activity distribution and ``infinite" MCGPU-PET simulation time, the addition of the 5D LUT to the simulator logic resulted in near-identical reconstructions to a reconstruction made from true coincidences, and the least object dependence, for all three phantoms. If a LUT is not included, the scatter sensitivity sinogram will be object dependent and lead to reconstruction artifacts.
    \item In the case where the activity distribution was known but MCGPU-PET simulation time was limited, we could get similar reconstruction accuracy to scenario 1 by also incorporating noise reduction for the scatter sinograms.
    \item In the case where the activity distribution was unknown and simulation time was limited, we included the LUT and noise reduction in an iterative scatter estimation procedure (i.e. joint estimation of the activity distribution and scatter sinogram), and still obtained accurate reconstructions.
\end{enumerate}

After the simulated evaluation, we adapted our approach and applied it to real data. The MC-based scatter estimation approach could produce scatter sinograms and reconstructions which were very close to those generated using the vendor implementation of tail-fitted SSS.